\documentclass[aps,pra,reprint,groupedaddress,amsmath,amssymb,preprintnumbers,showpacs,longbibliography]{revtex4-2}
\usepackage{graphicx}
\usepackage{hyperref}
\usepackage{amsmath,bm}
\usepackage{dcolumn}
\usepackage{braket}
\usepackage{physics}
\usepackage{float}
\usepackage{xcolor}
\newcommand{\eq}[1]{Eq.~(\ref{#1})}

\newcommand{\be}{\begin{equation}}
	\newcommand{\ee}{\end{equation}}
\newcommand{\ba}{\begin{eqnarray}}
	\newcommand{\ea}{\end{eqnarray}}
\newcommand{\bs}{\begin{subequations}}
	\newcommand{\es}{\end{subequations}}
\newcommand{\bw}{\begin{widetext}}
	\newcommand{\ew}{\end{widetext}}

\usepackage[final]{showlabels}
\usepackage{cleveref}
\crefname{equation}{Eq.}{Eqs.}
\crefname{figure}{Fig.}{Figs.}
\Crefname{Figure}{Fig.}{Figs.}
\crefname{table}{Table}{Tables}
\crefname{section}{Sec.}{Secs.}
\Crefname{section}{Sections}{Sections}
\crefname{subsection}{subsection}{subsections}
\Crefname{subsection}{Subsection}{Subsections}
\crefname{subsubsection}{subsection}{subsections}
\Crefname{subsubsection}{Subsection}{Subsections}
\crefname{paragraph}{subsection}{subsections}
\Crefname{paragraph}{Subsection}{Subsections}

\mathcode`\*="8000
{\catcode`\*\active\gdef*{\cdot}}

\begin{document}
\title{Correlation effects in high-harmonic generation from finite systems}
\date{\today}
\author{Thomas Hansen, Simon Vendelbo Bylling Jensen, and Lars Bojer Madsen}
\affiliation{Department of Physics and Astronomy, Aarhus
University, DK-8000 Aarhus C, Denmark}

\begin{abstract}
Using the Hubbard model we study how the process of high-order harmonic generation (HHG) is modified by beyond mean-field electron-electron correlation for both finite and bulk systems. A finite-size enhancement of the HHG signal is found and attributed to electrons backscattering off the lattice edges. Additionally, with increasing strength of the electron-electron correlation an enhancement of the high-frequency regime of the HHG spectrum is found. This is attributed to the on-site Coulomb repulsion between electrons giving rise to a localized quiver motion of the electrons. The finite-lattice enhancement dominates the HHG spectra from a few harmonic orders until a threshold from which the correlational enhancement dominates. This threshold is determined by the degree of correlation and decreases into the low-frequency regime for increasing electron-electron correlation. This infers that as the Mott insulator limit of high electron-electron correlation is approached, the finite-size effect on the electron dynamics becomes negligible. 
\end{abstract}
\maketitle

\section{Introduction}
For several decades high harmonic generation (HHG) has garnered much attention, mainly because of its potential for (i) spectrographic measurements of dynamics in the HHG sample on an ultrafast timescale, see, e.g., Refs.~\cite{Luu2018,Silva2018,Lein2002,Torres2007,Li2008} and (ii) production of coherent ultrashort ultraviolet laser pulses required for time-resolved measurements at the natural time-scale of electrons, see, e.g., Refs.~\cite{Lein2003,Schubert2014,Li2008}. A variety of models and approaches have been applied to describe HHG. For HHG in gases the prominent three-step model was developed, consisting of ionization, propagation and recombination \cite{Corkum1993,Lewenstein1994}. Ten years ago, HHG from solids was reported to exhibit different scaling behavior and different structure of the emitted spectrum as compared to atoms \cite{Ghimire2011}. This has led to increased interest in the field of HHG from solids, see Refs.~\cite{Vampa2017,Kruchinin2018,Ghimire2019,Yue2021} for recent reviews.

Several models have been used to study HHG from solids. For early studies, a model of choice was the semiconductor Bloch equations, which succeeded in separating the generation process into two coupled mechanisms \cite{Golde2008,Vampa2014}. These are inter- and intraband harmonic generation. Interband generation is similar to the three-step model of gaseous HHG, however, it is formulated in momentum-space and consists of (i) excitation of an electron from a given band to another, (ii) propagation of the electron and generated hole in their respective bands, and (iii) the recombination of the electron and hole. The intraband generation comes about from the propagation of electrons and holes through the individual bands of the bandstructure. Through the years, for experimental comparison, the semiconductor Bloch equations \cite{Schubert2014,Garg2016} and the semiclassical model for electron dynamics \cite{Luu2015,You2017,Ghimire2011,Kaneshima2018,Liu2017,Luu2018}, have been used. Recent modeling for extraction of potentials and densities from HHG experiments in solids have used an accelerated frame picture \cite{Lakhotia2020}, where the harmonics are generated as the laser-driven density probes the force from the potential \cite{Madsen2021}. Most applications of these approaches rely on the independent-electron approximation and a long-range periodicity of the solid, as they consider electrons propagating through a bandstructure. Going beyond this are methods based on time-dependent density functional theory (TDDFT) for HHG, which, in principle, include beyond mean-field electron-electron correlations \cite{Yamada2021,PhysRevA.96.053418,PhysRevLett.118.087403,Floss2018}. TDDFT is not restricted on the periodicity of the solid and has thus been used to study HHG in, e.g., finite systems \cite{Hansen2018} and systems with topological edgestates \cite{Hansen2018_2}.

Recently, it was found that an enhancement of the HHG spectrum arises in finite lattices due to electrons backscattering on the edges of the lattice \cite{Chuan2021}. In the finite-size limit, however, electron-electron correlations are of importance as alluded to in HHG experiments with monolayer materials \cite{Liu2017} and predicted for pump-probe HHG \cite{Jensen2021,Jensen2021_2}. To investigate beyond mean-field electron-electron correlations in HHG, several effective Hamiltonian models have been developed and applied \cite{PhysRevLett.124.157404,Murakami2018,Silva2018,Takayoshi2019,PhysRevLett.121.097402,Murakami2021,Lysne2020}. A prominent example of such methods is the Hubbard model, which predicted that increasing beyond mean-field electron-electron correlation for a periodic system would lead to enhancement in the HHG spectra \cite{Silva2018} and increasing cutoff energy \cite{Murakami2018}. This model has led to a link between enhancement of the HHG spectrum and doublon-holon pair dynamics in the lattice, resulting in a thee-step model-like picture of HHG from correlation effects \cite{Murakami2018,Murakami2021}.

In this work we present a combined study of finite-system effects and beyond mean-field correlation effects of HHG. We do so by considering predictions from the Hubbard model for both finite and bulk systems. In doing so we aim to answer the following, (i) How is the HHG spectrum modified by the interplay of finite system effects and beyond mean-field correlation effects? (ii)  What electron dynamics can be attributed to the modifications of the HHG spectra?
 
The paper is organized as follows. In Sec.~II, the model and simulation methods are presented. In Sec.~III, the results are presented and discussed, and in Sec.~IV we conclude. Atomic units are used throughout unless stated otherwise.

\section{Theoretical model and methods}\label{sec:theory}

The field-free Hamiltonian for a system of electrons moving on a lattice can be written as \cite{Hubbard}
\begin{align}
\hat{H}_{\mathrm{Hub}}&= \sum_{i,j}\sum_\sigma t_{ij}\hat{c}_{i,\sigma}^\dagger \hat{c}_{j,\sigma}\nonumber\\
&+\sum_{i,j,k,l}\sum_{\sigma,\gamma} U_{ijkl}\hat{c}_{i,\sigma}^\dagger\hat{c}_{j,\gamma}^\dagger \hat{c}_{k,\gamma} \hat{c}_{l,\sigma},
\end{align}
where $\hat{c}_{i,\sigma}^\dagger$ and $\hat{c}_{i,\sigma}$ are, respectively, the fermionic creation and annihilation operators of an electron on site $i$ with spin $\sigma = \lbrace \uparrow, \downarrow \rbrace$. The hopping term $t_{ij}$ accounts for electron hopping from site $j$ to site $i$ including the mean-field part of the electron-electron interaction. Further, $U_{ijkl}$ accounts for the beyond mean-field electron-electron interaction between two electrons starting on site $k,l$ and ending on sites $j,i$, respectively. In this work, we apply the on-site approximation to the correlation term and the nearest-neighbor approximation to the hopping term thus denoting it NN. The Hamiltonian thus reduces to
\begin{align}
	\hat{H}_{\mathrm{NN}}&=\sum_{i,\sigma}\left(t_{ii+1}\hat{c}_{i,\sigma}^\dagger \hat{c}_{i+1,\sigma}+ \mathrm{h.c.}\right)\nonumber\\
	&+\sum_{i,\sigma}t_{ii}\hat{n}_{i,\sigma}+\sum_{i}\left(U_{iiii}\hat{n}_{i,\uparrow} \hat{n}_{i,\downarrow}\right), \label{eq:t_ii}
\end{align}
with $\hat{n}_{i,\sigma} = \hat{c}_{i,\sigma}^\dagger \hat{c}_{i,\sigma}$ being the number operator of electrons on site $i$ with spin $\sigma$. We consider half-filling, which means the $t_{ii}$ term supplies a constant energy-shift, which we set to zero by shifting our energy scale. Further, the beyond mean-field electron-electron interaction is set to be constant and independent of the lattice site, i.e., $U_{iiii}=U$. Finally we express the hopping term as $t_{i,i+1}=-t_0\mathrm{e}^{\mathrm{i}aA(t)}$. Here the exponential phase is Peierl's phase, which is used to include the interaction with an electromagnetic field, within the dipole approximation, expressed through the vector potential $A(t)$ and the lattice spacing $a$ \cite{Hubbard}. We apply the nearest-neighbor hopping term $t_0=0.0191$ a.u., and a lattice spacing $a=7.56$ a.u. Both parameters are chosen to mimic the values from $\mathrm{Sr}_2\mathrm{CuO}_3$ \cite{Tomita2001} as in Ref.~\cite{Silva2018}. $U$ will be treated as a parameter and always given in units of $t_0$. The applied Hubbard Hamiltonian  including interaction with the external field is thereby given as  \cite{Hubbard}

\begin{align}
\hat{H}&=-t_0\sum_{i,\sigma}\left(\mathrm{e}^{\mathrm{i} a A(t)}\hat{c}_{i,\sigma}^\dagger \hat{c}_{i+1,\sigma}+ \mathrm{h.c.}\right)\nonumber\\
&+U\sum_{i}\left(\hat{n}_{i,\uparrow} \hat{n}_{i,\downarrow}\right). \label{eq:Hamilton incl A}
\end{align}

The applied $N_c=10$ cycle electromagnetic field is linearly polarized along the lattice dimension and defined from the vector potential 
\begin{align}
A(t)&= A_0 \cos(\omega_L t-N_c\pi)\sin^2\left(\frac{\omega_L t}{2*N_c}\right).
\end{align}
The amplitude $A_0 = F_0 /\omega_L$ is defined from the peak electric field-strength $F_0=0.97\times10^{-3}$ a.u., corresponding to a peak intensity of $3.3\times10^{10}$ W/cm$^2$ and the laser carrier frequency is chosen as $\omega_L=0.005$ a.u. corresponding to a $33$ THz field.

We consider simulation of finite and bulk lattices. In both cases a lattice consisting of $L=12$ sites is used. Bulk lattices can be approximated by applying periodic boundary conditions to the system. Finite lattices will be simulated by applying closed boundary conditions, i.e., there are no states outside of the one-dimensional chain. In other words, for the finite system, we neglect the coupling to the vacuum. This procedure is justified by two conditions (i) The lattice length is more than two times the quiver radius of the free-electron motion in the electromagnetic field, $a*L>2\frac{F_0}{\omega^2}$ for the considered laser parameters. Classically, this requirement infers that the majority of a uniform electron distribution would not propagate to the surrounding vacuum \cite{Hansen2018}. (ii) An insignificant part of the electron population is transferred to states with energy above the ionization energy. This requirement can be fulfilled by requiring that an insignificant electron population is transferred to states above the zero-point energy of \eq{eq:Hamilton incl A}. The energy of the ground state of the Hamiltonian of \eq{eq:Hamilton incl A} is increasing with increasing $U$. The highest $U$ considered in this report, $U=0.8 t_0$, will thus have the least bound ground state. For the $U$-value of $0.8 t_0$ the ground state energy is negative with an absolute value of $\approx 47$ orders of the laser-carrier frequency. At such harmonic orders the calculations below show that the HHG spectrum has decreased by a factor $\approx 10^{13}$, compared to its peak. Thus, transitions to states with energy above the vacuum energy can safely be neglected. 

The time-dependent Schr\"{o}dinger equation is propagated using the Arnoldi-Lancoz algorithm, for both the imaginary and real time propagation. The ground state of the Hamiltonian, \eq{eq:Hamilton incl A}, is employed as the initial state of the simulation. It is found by imaginary time propagation. Three separate convergence criteria are used for the ground state: (i) the energy should have converged, (ii) the state should respect the lattice symmetry, i.e., the ground state should be inversion (translation) symmetric for the finite (bulk) system, respectively and (iii) the number of doublon-holon pairs, i.e., the measure of \eq{eq:D-measure} below, should have converged. 

The spectrum,  $S(\omega)$, is calculated based on the time derivative of the current as
\begin{align}
S(\omega)&=\left|\mathcal{F}\left(\dv{t}j(t)\right)\right|^2=\left|\omega j\left(\omega\right)\right|^2, 
\label{eq:spectra}
\end{align}
with $j(\omega)$ being the Fourier transform ($\mathcal{F}$) of the nearest-neighbor current $j(t)$, which is found from \cite{Mahan}
\begin{align}
j(t)&=\left\langle \hat{j}(t)\right\rangle=\sum_{i=1}^{N_j}\sum_{\sigma=\left\{\uparrow,\downarrow\right\}}  j_{i, \sigma}(t), \label{eq:N_j}
\end{align}
where $j_{i,\sigma}(t)=\left\langle \hat{j}_{i,\sigma}(t)\right\rangle$ and $N_j$ is the number of sites for which a transition from site $i$ to $i+1$ is possible, i.e. $N_j=12$ for the bulk lattice and $N_j=11$ for the finite lattice, due to the different boundary conditions.  In the equation just below \eq{eq:N_j}, we let $\hat{j}_{i,\sigma}$ denote the site- and spin-dependent current operator \cite{Hubbard}
\begin{align}
\hat{j}_{i, \sigma}(t) &=\mathrm{i} a t_{0}\left[\mathrm{e}^{\mathrm{i} a A(t)} \hat{c}_{i, \sigma}^{\dagger} \hat{c}_{i+1, \sigma}-\text { h.c. }\right].
\label{eq:current operator}
\end{align}
Note that each $j_{i,\sigma}(t)$ may be interpreted as the net current between sites $i$ and $i+1$. This is found from the current flowing from site $i+1$ to site $i$ subtracted with the current flowing from site $i$ to site $i+1$.
	
In the analysis of our results, we furthermore consider the expectation value of the number of doublon-holon pairs, which will be used for analysis. We denote this measure as the $D$-measure
\begin{align}
D&= \frac{1}{L}\sum_{i=1}^{L}\left\langle\hat{n}_{i,\uparrow}\hat{n}_{i,\downarrow}\right\rangle. \label{eq:D-measure}
\end{align}
Note that, as half-filling is used, the number of doublons will always equal the number of holons. That leads to $0\leq D\leq 1/2$.

\section{Results}

\subsection{HHG spectra for different values of $U$}
\begin{figure}
\includegraphics[width=\linewidth]{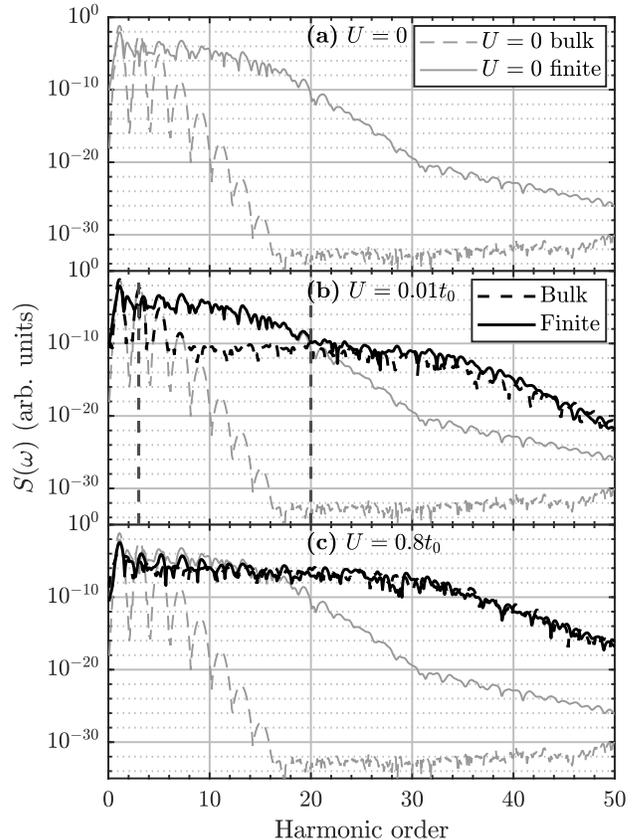}
\caption{HHG spectra [\cref{eq:spectra}] from bulk and finite lattices with (a) $U = 0$, (b) $U = 0.01 t_0$ and (c) $U = 0.8 t_0$. The spectra with $U=0$ from (a) are also given in (b) and (c) for direct comparison. See text for laser and system parameters. The vertical dashed lines in (b) indicate the splitting into three characteristic regions as discussed in Sec.~III B.}
\label{fig:spectra U comp}
\end{figure}

In order to illustrate the effects of including beyond-mean-field electron-electron interaction three representative $U$-values have been chosen $U=0$, $U=0.01 t_0$, and $U=0.8 t_0$. These values depict the characteristic behaviors found from an extensive scan of $U$-values.  Figure \ref{fig:spectra U comp} depicts the HHG spectra from bulk and finite lattices for these $U$ values. For uncorrelated electrons, ($U=0$), \cref{fig:spectra U comp}(a), an enhancement is found for the finite system, similar to the results of Ref.~\cite{Chuan2021} where an uncorrelated chain was considered within a TDDFT approach.  This enhancement peaks at the 16' to 20'th harmonic order, with an enhancement by a factor of $10^{20}-10^{25}$. This feature persists in the regime of weak electron-electron correlation of \cref{fig:spectra U comp}(b) but has decreased to a factor $\approx 10^{5}$ at its peak, between the 8'th and 16'th harmonic order.

Identifying the regions in which electron-electron correlation effects and finite lattice effects are most critical to the spectra is of interest. This identification can be done by comparing a given non-zero $U$ finite lattice spectrum to a key set of other spectra. As an example of this, we consider the $U=0.01 t_0$ finite lattice spectrum from \cref{fig:spectra U comp}(b). Comparing the $U=0.01 t_0$ finite lattice spectrum (black, full) to the $U=0$ bulk lattice spectrum (grey, dashed) yields that neither electron-electron correlation nor finite lattice effects are important for the gain in the 1st and 3rd harmonic orders. This is seen by the high degree of overlap between the two spectra in this region. 
Next, if comparing the $U=0.01 t_0$ finite and bulk lattice spectrum (black, full and black, dashed), a finite size enhancement is found from the 5'th till approximately the 20'th harmonic order. The reason for attributing this difference to finite size enhancement and not correlation is that in this region the $U=0.01 t_0$ finite lattice spectrum is identical to the $U=0$ finite lattice spectrum at which solely finite lattice effects but no electron-electron correlation effects are present.
For the region after the 20'th harmonic order, electron-electron correlation effects dominate for both the $U=0.01 t_0$ finite and bulk lattice spectra, as these are converging to a spectrum which does not resemble any of the $U=0$ spectra. 
Similar comparisons between the $U=0.8 t_0$ finite and bulk lattice spectra of  \cref{fig:spectra U comp}(c) with the $U=0$ finite and bulk lattice spectra yield another conclusion. Here, in the regime of relatively high $U$-values, the finite lattice effects never dominate, seen by the lack of overlap between the $U=0.8 t_0$ finite lattice spectrum and the $U=0$ finite lattice spectrum and the large resemblence between the $U=0.8 t_0$ finite and bulk lattice spectra. If considering the first to fifth harmonic orders in  \cref{fig:spectra U comp} (c), electron-electron correlation effects result in a decrease in the spectrum compared to the $U=0$ case. The reason for this may be that such low order harmonics are generated from trajectories spanning multiple lattice sites, which, as discussed below,  become decreasingly likely with increasing correlation, as beyond mean-field interactions occur across the entire lattice.

\subsection{Three characteristic spectral regions}
In general, the spectra can be split into three regions. For low electron-electron correlation, \cref{fig:spectra U comp}(b), the first regime of the first couple of harmonics are unaffected by electron-electron correlation and finite lattice effects. Hereafter in the second, middle region the finite lattice enhancement dominates. For $U=0.01 t_0$ this middle region is from the 5'th to 20'th harmonic order. The high-frequency limit of this region is decreasing with increasing $U$.  As discussed below, this second region is linked to electron trajectories which are modified by electron-edge interactions. Such trajectories likely span several lattice sites as the electrons need to travel to interact with the edge. Finally, the high-frequency region is dominated by electron-electron correlation effects. The harmonics in this region are linked to high-frequency oscillatory electron trajectories. Such quiver-like trajectories are possible for all electrons across the entire lattice which are all affected by electron-electron correlations. For increasing $U$-values the size of the middle region decreases and vanishes, and the entire spectrum eventually becomes dominated by the electron-electron correlation. Increasing $U$ to $U=0.8t_0$ also affects the first region of low harmonic orders, where a decrease is found in the spectra, but from around the 5'th order and upwards a gain in the spectrum is seen. The three regions are indicated in \cref{fig:spectra U comp} (b) by vertical dashed lines. 

\subsection{Population dynamics analysis}
\begin{figure}
\includegraphics[width=\linewidth]{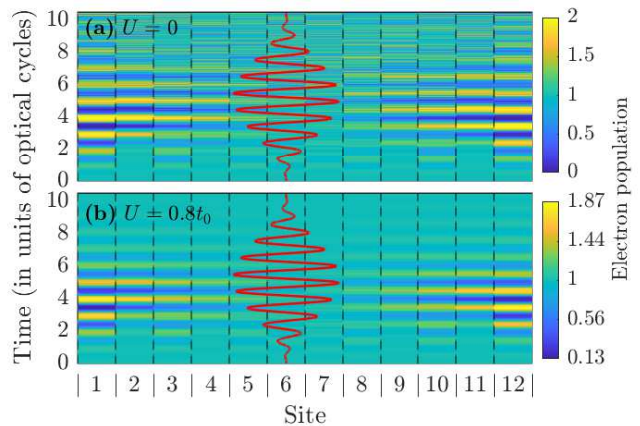}
\caption{The electron population on the various sites in the finite lattice over time for (a) $U=0$  and (b) $U=0.8 t_0$. Plotted on top in red is the driving electric field. Note that the colorbars are different in (a) and (b).}
\label{fig:popst}
\end{figure}

To achieve insight into the dynamics at hand, the evolution of the real-space, i.e., the site-specific, population during the driving pulse is considered in \cref{fig:popst}. The population at the various sites of the finite lattice for $U=0$ and $U=0.8 t_0$ is shown, together with the electric field. The $U=0.01 t_0$ population has been omitted, as the resulting plot is indistinguishable from the $U=0$ case. The $U=0.01 t_0$ being indistinguishable from the $U=0$ plots is the case for all plots considered in the analysis of the dynamics. Therefore it suffices to consider the cases $U=0$ and $U=0.8t_0$ in 
\cref{fig:popst,fig:D,fig:current,fig:current Jsplit}. The bulk lattice cases are not shown in \cref{fig:popst}, since the population on every site remains constant throughout the entire simulation, due to the translational symmetry of the system.

In \cref{fig:popst}, it is observed that the electron population is oscillating forwards and backwards in the lattice with backscattering on the lattice edges, in a similar manner as the oscillations of a free electron under the plotted electric field. Furthermore, as $U$ increases, the magnitude of the population at the edges of the lattice decreases, as seen by comparing the populations in Figs. \ref{fig:popst} (a) and (b). This decrease is due to the increased energy associated with high electron population on a given site. Note that as a result of Pauli's exclusion principle, the electrons cannot all reach the edge sites, but end up piling up further from the edge. This allow the electrons to backscatter with increasing probability without reaching all the way to the lattice edge. In the $U=0$ case, the population is close to two for both sites closest to the lattice edge, shortening the effective length of the chain in which the remaining electrons can oscillate. The electron build-up at the system edges at certain times does not entail that coupling to vacuum states should be included, as the previous arguments of \cref{sec:theory} regarding the negligible coupling to vacuum still hold.
	
\subsection{Doublon-holon pair analysis}
\begin{figure}
\centering
\includegraphics[width=\linewidth]{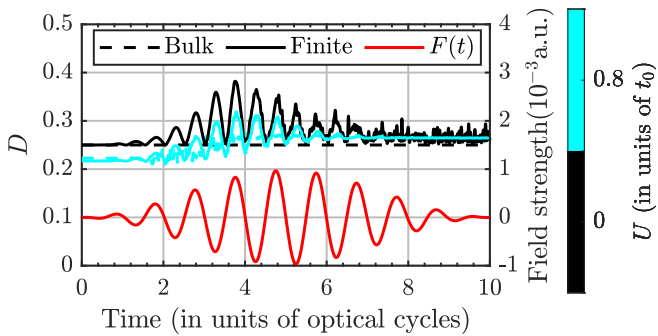}
\caption{The $D$-measure [\cref{eq:D-measure}] for both finite and bulk lattices with $U=0$ and $U=0.8 t_0$. The applied electric field is plotted for comparison in the lower part of the figure. }
\label{fig:D}
\end{figure}
Another way to illustrate some of the differences between the dynamics of the finite and bulk systems is through the $D$-measure of \eq{eq:D-measure}. As is clear from \eq{eq:D-measure}, the term $\left\langle \hat{n}_{i, \uparrow} \hat{n}_{i,\downarrow} \right\rangle$  returns unity if a site is doubly occupied. The corresponding quasiparticle is denoted a doublon, similarly if a site is unoccupied the quasiparticle is denoted a holon. As $U$ increases, double occupancy is less likely and $D$ decreases. In \cref{fig:D}, the $D$-measure with $U=0$ and $U=0.8 t_0$, for both bulk and finite lattice is shown alongside the electric field. Initially the $D$-measure of the ground state in the finite and bulk lattice are identical for $U=0$ and takes the value $D=0.25$. This similarity is likely a result of the population in the basis states being independent of the number of doublon-holon pairs in said basis state, as $U=0$. For higher $U$, the finite lattice ground state has a lower value of $D$ compared to the bulk lattice ground state. This trend is consistent with a variety of other $U$-values (not shown) and the difference is due to the boundary conditions on the lattice. The extra coupling in the bulk lattice causes it to be energetically favorable for the electrons to distribute themselves more evenly across the basis states, in accordance with the $U=0$ ground-state. This causes the $U\neq 0$ ground state $D$-value of the bulk lattices to be closer to the $D=0.25$ of the $U=0$ cases than the finite lattice cases. The bulk $U=0$ simulation shows no change in $D$ as a function of time. This is a result of there being no energetical cost or gain to the creation of doublon-holon pairs for $U=0$. The $U=0.8 t_0$ bulk system starts at a lower $D$-value than the $U=0$ case as is to be expected from the increased electron-electron repulsion. As the system is excited by the pulse, higher energy states are populated and $D$ increases. The finite $U=0$ case shows a series of peaks at twice the pulse frequency. These peaks occur at the same instants of time as the edge-states form [see \cref{fig:popst}], and are a direct consequence thereof. The $U=0.8 t_0$ finite case shows a similar overall increase in $D$ as the $U=0.8 t_0$ bulk case and a peak structure similar to that of the $U=0$ finite lattice case. The height of the peaks is smaller than in the $U=0$ finite case, due to the energy associated with high $D$.

\subsection{Current analysis}
\begin{figure}
\centering
\includegraphics[width=\linewidth]{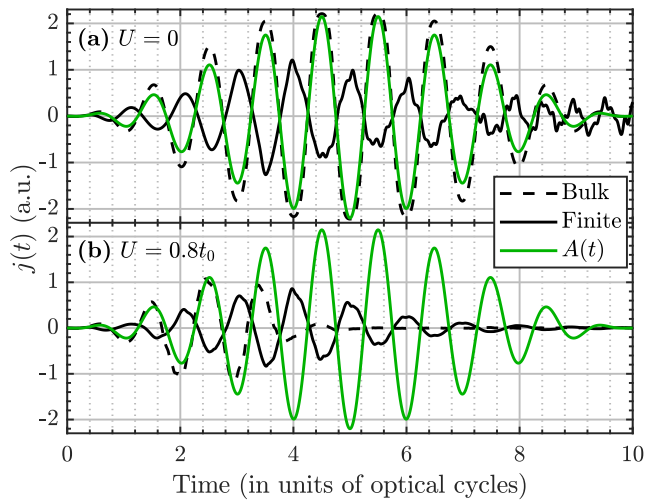}
\caption{The current in both the finite and bulk lattice as a function of time for (a) $U=0$, (b) $U=0.8 t_0$, and (c) $U=2 t_0$. The green curve shows the applied electromagnetic vector potential $A(t)$.}
\label{fig:current}
\end{figure}

The current relates the electron dynamics to the spectra [\eq{eq:spectra}]. Figure \ref{fig:current} shows the current for bulk and finite lattices with $U=0$ and $U=0.8 t_0$, as well as the electromagnetic vector potential of the pulse for comparison. For $U=0$ the Hubbard model reduces to the tight-binding model. The one-dimensional, one-band tight-binding model results in a band structure given as \cite{Solid_State_Physics}
\begin{align}
	\epsilon(k)&=E_0-2t_0\cos(k*a),
\end{align}
 where $\epsilon(k)$ is the band-energy at crystal-momentum $k$ and $E_0$ is an energy shift. It is well-known that the  intraband current from a $k$-space localized wavepacket is given by $j(t)\propto \left. \frac{\mathrm{d}\epsilon(k)}{\mathrm{d}k}\right|_{k=k(t)}$  and semiclassically $k(t)=k(0)+A(t)$. In this case, as $j(t=0)=0$, $k(0)=0$, the current reads
 \begin{align}
 j(t)&\propto -\sin\left(a*A(t)\right).
 \end{align}
 This result for the current is in exact agreement with the $U=0$ bulk lattice result of \cref{fig:current} (a) but of course not for the finite lattice case as $k$ is only a good quantum number for infinite periodic lattices.

 The $U=0$ finite lattice case, also seen in \cref{fig:current}(a), is out of phase with the vector potential. This is a consequence of the electrons being driven to the edge of the lattice. At the edge, the acceleration on the electrons from the electric field is being balanced by the acceleration away from the edge-state induced by the system boundary. Once the electric field decreases, the electrons start moving out of the edge-state. The electrons are thus moving in the direction opposite to the field markedly sooner than in the bulk case, where the electrons need be decelerated first. 
 
Figure \ref{fig:current} (b) shows, when compared with \cref{fig:current} (a), that the overall magnitude of the current decreases with increasing $U$. This is consistent with the spectra since the 1'st and 3'rd harmonics, which dominate the spectra, decrease in signal strength with increasing $U$. The finite currents from both Figs.~\ref{fig:current}(a) and \ref{fig:current}(b) show sharper and sharper peaks for times up till the 4'th optical cycle. A probable explanation is that the electrons moving across the lattice at increasing speed cause the shift from acceleration away from an edge-state to deceleration into an edge-state to happen quicker and quicker, resulting in sharper and sharper peaks in the current. 
	In all cases where beyond mean-field electron-electron correlation effects are included, i.e., $U\neq0$, the currents start decreasing at some point before the peak of the driving pulse. This happens for the finite cases at around the fourth optical cycle, in the $U=0.8 t_0$ bulk lattice case after the third optical cycle. Scattering cause the electron dynamics to become more disordered resulting in oscillations at low frequencies becoming unlikely as time progresses.  By a Gabor analysis we find that, similarly to the low-order harmonics, all regimes of the spectra are generated dominantly around the peak of the current. It seems that the finite lattice keeps the electron dynamics orderly longer than the bulk lattice for $U\neq 0$. This may be because the oscillation from edge-state to edge-state helps keep the electrons moving in a coordinated manner. 
\begin{figure}
\centering
\includegraphics[width=\linewidth]{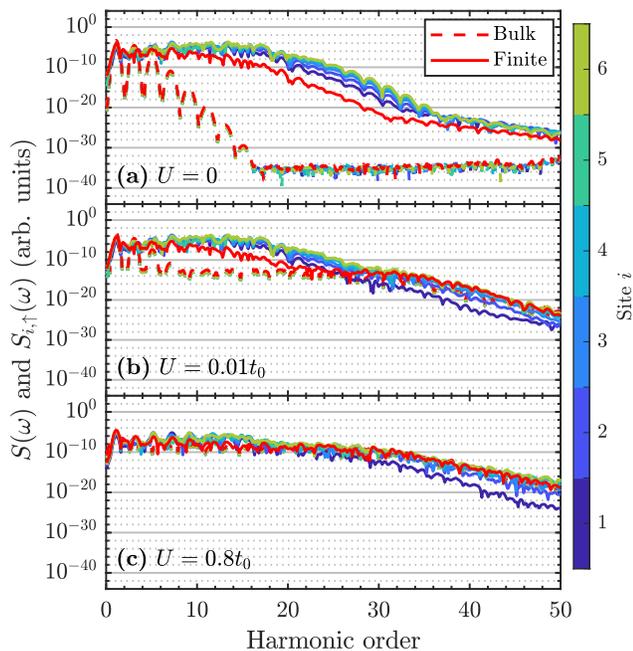}
\caption{HHG spectra generated from the currents between individual sites, see \cref{eq:S_i_sigma}, for bulk and finite lattices and three $U$-values. The spectra of the total current [\cref{eq:spectra}] is divided by $4N_{j}^2$ and ploted by full and dashed red curves.}
\label{fig:spectra Jsplit}
\end{figure}

\subsection{Site-specific currents and spectra}
The system can be further examined by considering the currents associated with the individual sites, that is the $j_{i,\sigma}(t)$'s of \cref{eq:N_j}. Here, the lattice of $L=12$ sites will generate $2N_{j}$ different $j_{i,\sigma}(t)$'s, see \cref{eq:N_j}.
The currents for both spins are indistinguishable. Therefore only spin-up electrons are considered for the present analysis. Further, the lattice is symmetrical around the lattice center. So rather than displaying all $j_{i,\sigma}$'s it is sufficient to display $L/2=6$. In doing so, the terms are numbered from the edge, so that $j_{1,\sigma}(t)$ is the current between site $1$ and $2$, which is identical to that of site $11$ and $12$, both located at the edge of the lattice. Similarly $j_{6,\sigma}(t)$ is the current between site $6$ and $7$ at the center of the lattice. For the corresponding bulk case, the current between each site is identical, due to the translational symmetry.  Since the observed spectra can be expressed as [\cref{eq:spectra,eq:N_j,eq:current operator}]
\begin{align}
	S(\omega)&=\left|\sum_{i=1}^{N_j}\sum_{\sigma=\left\{\uparrow,\downarrow\right\}}\mathcal{F}\left(\frac{\mathrm{d}}{\mathrm{d}t}j_{i, \sigma}(t)\right)\right|^2,\nonumber\\
	&=\left|\sum_{i=1}^{N_j}\sum_{\sigma=\left\{\uparrow,\downarrow\right\}}\left(\omega j_{i, \sigma}(\omega)\right)\right|^2,
\end{align} 
 it may be of interest to inspect 
 \begin{align}
 	S_{i,\sigma}(\omega)=|\omega j_{i, \sigma}(\omega)|^2, \label{eq:S_i_sigma}
 \end{align}
i.e., the spectra generated from singling out the current between site $i$ and site $i+1$ of electrons with spin $\sigma$. By summing all $S_{i,\sigma}(\omega)$'s one does not find the total spectrum, as one would need to also account for interferences. However, $S_{i,\sigma}(\omega)$ spectra can reveal from which sites the dominant contributions to the total current arise, and shed light on any asymmetries in the electron dynamics throughout the lattice. Such spectra for (a) $U=0$ , (b) $U=0.01t_0$, and (c) $U=0.8t_0$ are shown in \cref{fig:spectra Jsplit} for both bulk and finite lattices.
 For comparison, the spectra generated from the total current, i.e., the spectra from \cref{fig:spectra U comp}, are given for each case, divided by $4N_{j}^2$ to compare on equal footing. In all bulk lattice cases, the spectra generated from the total current overlap with the $S_{i,\sigma}(\omega)$'s. This is in agreement with the symmetries of the bulk lattice. For $U=0$, in \cref{fig:spectra Jsplit} (a), it is seen that the spectra from the finite lattice is generated dominantly at the center of the lattice across a wide range of HH-frequencies. At the edges, the electrons change direction, due to the electron-edge scattering, contributing to a reduced average velocity reducing the current and spectrum. Oppositely, closer to the center of the lattice, the electrons are moving unaffected by the edges, leading to high current and spectrum. As $U$ increases, see Figs.\ref{fig:spectra Jsplit} (b) and (c), the differences between the $S_{i,\sigma}$'s diminish. This is similar to how the difference between the bulk and finite lattice spectra diminishes, as discussed in connection with \cref{fig:spectra U comp}. This behavior is attributed to more significant electron-electron interaction events making the effects of electron-edge scattering less significant overall for increasing $U$. 
	
\begin{figure}
\centering
\includegraphics[width=\linewidth]{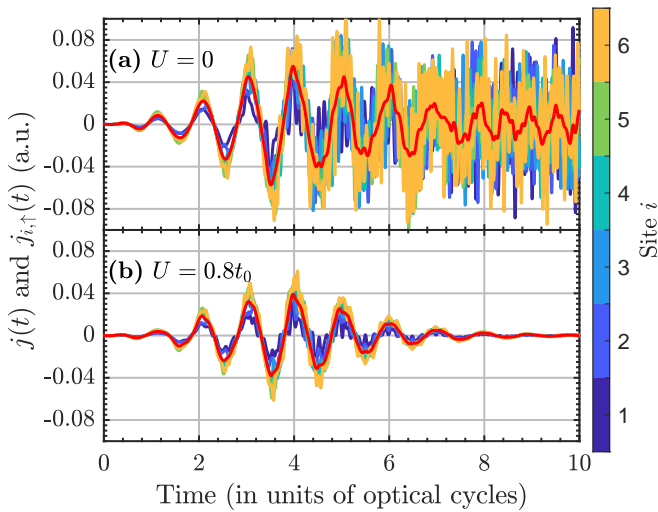}
\caption{Plots of the currents between the different lattice sites for the finite lattice and two different values of $U$. The total current is given in red when divided by $N_{j}$ [\cref{eq:N_j}].}
\label{fig:current Jsplit}
\end{figure}

The currents from the individual sites are investigated in \Cref{fig:current Jsplit}. The $j_{i,\sigma}(t)$'s for the finite case are shown with the same $U$-values as used for \cref{fig:current}, that is: $U=0$, and $U=0.8 t_0$, as well as the total current divided by $2N_j$. The bulk results have been omitted due to the current on all sites being identical. In \cref{fig:current Jsplit} (a) with $U=0$, the current across the individual sites are observed to be similar before the fourth optical cycle. Hereafter the $j_{i,\sigma}(t)$'s show wildly varying results, indicating much more disorderly electron dynamics, consistent with the earlier analysis. This is not the case in \cref{fig:current Jsplit} (b) with $U=0.8 t_0$, likely due to the increased number of scattering events across the lattice causing the dynamics between each pair of neighboring sites to become increasingly similar, and thus resulting in the scattering on the edges becoming a much less significant effect when compared to scattering between the electrons themselves. This would cause the current on each sites in the lattice to become similar, consistent with the results in \cref{fig:current Jsplit} (b). Note, however, that the high electron-electron correlation does not necessarily mean that the electron dynamics have become ordered, merely that scattering events are taking place at every site in the lattice, making the edge sites less distinct. One might expect the pulse to drive the electrons also after the peaking of the pulse. Here, however, the $U=0.8 t_0$ results are continually decreasing. As the electrons seem to oscillate at much higher frequency than the driving pulse in the later half of the pulse, the effect of the pulse on the electrons is likely averaged out.

\section{Summary and Conclusion}
In this work, we have used the laser-driven Hubbard model to examine the interplay between finite system effects and electron-electron correlation in HHG. Previously, an enhancement in finite systems was reported \cite{Chuan2021} when not accounting for beyond mean-field electron-electron correlation, i.e., the $U=0$ case of this study. The Hubbard model was used earlier to demonstrate beyond mean-field electron-electron correlation effects of HHG enhancement, leading to an enhancement of the high frequency regime \cite{Silva2018}. Combining consideration of both finite and correlation effects, we set out to address the two questions that we asked in the introduction regarding (i) the modification of the spectra due to the interplay between finite size and electron-electron correlation effects and (ii) an investigation of the underlying electron dynamics. We found a modification of the spectra in two separate regions for weak correlation. In the weak correlation case the first couple of harmonic orders, 1'st to 3'rd for $U=0.01t_0$, are largely unaffected by finite lattice and correlation effects. The following harmonic orders, around 5'th to 20'th for $U=0.01 t_0$, are heavilly affected by the finite lattice effects, resulting in a noticeable gain in the spectrum. Finally in the high-frequency regime, 20'th harmonic and up for $U=0.01t_0$, is dominated by correlation effects, also resulting in a gain in the spectrum. For stronger correlation, see the $U=0.8t_0$ finite lattice result in \cref{fig:spectra U comp} (c), the spectrum is heavily affected by correlation effects throughout the entire spectrum. 
	
The underlying dynamics are different for each of the three regions found in the weak correlation limit. The first couple of harmonic orders are associated with trajectories spanning multiple lattice sites largely unaffected by finite-lattice and correlation effects. The second region, 5'th to 20'th region for $U=0.1t_0$, is associated with trajectories, which also span multiple lattice sites, but which backscatter on the lattice edges resulting in quick momentum changes, associated with higher harmonic orders. Finally, the highest orders are associated with high-frequency quiver-like trajectories. Such trajectories are possible as a result of correlation interactions taking place throughout the lattice. 

\acknowledgments
This work was supported by the Independent Research Fund Denmark (Grant No. 9040-00001B and 1026-00040B).


\begin{thebibliography}{42}%
\makeatletter
\providecommand \@ifxundefined [1]{%
 \@ifx{#1\undefined}
}%
\providecommand \@ifnum [1]{%
 \ifnum #1\expandafter \@firstoftwo
 \else \expandafter \@secondoftwo
 \fi
}%
\providecommand \@ifx [1]{%
 \ifx #1\expandafter \@firstoftwo
 \else \expandafter \@secondoftwo
 \fi
}%
\providecommand \natexlab [1]{#1}%
\providecommand \enquote  [1]{``#1''}%
\providecommand \bibnamefont  [1]{#1}%
\providecommand \bibfnamefont [1]{#1}%
\providecommand \citenamefont [1]{#1}%
\providecommand \href@noop [0]{\@secondoftwo}%
\providecommand \href [0]{\begingroup \@sanitize@url \@href}%
\providecommand \@href[1]{\@@startlink{#1}\@@href}%
\providecommand \@@href[1]{\endgroup#1\@@endlink}%
\providecommand \@sanitize@url [0]{\catcode `\\12\catcode `\$12\catcode
  `\&12\catcode `\#12\catcode `\^12\catcode `\_12\catcode `\%12\relax}%
\providecommand \@@startlink[1]{}%
\providecommand \@@endlink[0]{}%
\providecommand \url  [0]{\begingroup\@sanitize@url \@url }%
\providecommand \@url [1]{\endgroup\@href {#1}{\urlprefix }}%
\providecommand \urlprefix  [0]{URL }%
\providecommand \Eprint [0]{\href }%
\providecommand \doibase [0]{https://doi.org/}%
\providecommand \selectlanguage [0]{\@gobble}%
\providecommand \bibinfo  [0]{\@secondoftwo}%
\providecommand \bibfield  [0]{\@secondoftwo}%
\providecommand \translation [1]{[#1]}%
\providecommand \BibitemOpen [0]{}%
\providecommand \bibitemStop [0]{}%
\providecommand \bibitemNoStop [0]{.\EOS\space}%
\providecommand \EOS [0]{\spacefactor3000\relax}%
\providecommand \BibitemShut  [1]{\csname bibitem#1\endcsname}%
\let\auto@bib@innerbib\@empty
\bibitem [{\citenamefont {Luu}\ and\ \citenamefont
  {W{\"o}rner}(2018)}]{Luu2018}%
  \BibitemOpen
  \bibfield  {author} {\bibinfo {author} {\bibfnamefont {T.~T.}\ \bibnamefont
  {Luu}}\ and\ \bibinfo {author} {\bibfnamefont {H.~J.}\ \bibnamefont
  {W{\"o}rner}},\ }\bibfield  {title} {\bibinfo {title} {Measurement of the
  {{B}}erry curvature of solids using high-harmonic spectroscopy},\ }\href
  {https://doi.org/10.1038/s41467-018-03397-4} {\bibfield  {journal} {\bibinfo
  {journal} {Nature Communications}\ }\textbf {\bibinfo {volume} {9}},\
  \bibinfo {pages} {916} (\bibinfo {year} {2018})}\BibitemShut {NoStop}%
\bibitem [{\citenamefont {Silva}\ \emph {et~al.}(2018)\citenamefont {Silva},
  \citenamefont {Blinov}, \citenamefont {Rubtsov}, \citenamefont {Smirnova},\
  and\ \citenamefont {Ivanov}}]{Silva2018}%
  \BibitemOpen
  \bibfield  {author} {\bibinfo {author} {\bibfnamefont {R.~E.~F.}\
  \bibnamefont {Silva}}, \bibinfo {author} {\bibfnamefont {I.~V.}\ \bibnamefont
  {Blinov}}, \bibinfo {author} {\bibfnamefont {A.~N.}\ \bibnamefont {Rubtsov}},
  \bibinfo {author} {\bibfnamefont {O.}~\bibnamefont {Smirnova}},\ and\
  \bibinfo {author} {\bibfnamefont {M.}~\bibnamefont {Ivanov}},\ }\bibfield
  {title} {\bibinfo {title} {High-harmonic spectroscopy of ultrafast many-body
  dynamics in strongly correlated systems},\ }\href
  {https://doi.org/10.1038/s41566-018-0129-0} {\bibfield  {journal} {\bibinfo
  {journal} {Nature Photonics}\ }\textbf {\bibinfo {volume} {12}},\ \bibinfo
  {pages} {266} (\bibinfo {year} {2018})}\BibitemShut {NoStop}%
\bibitem [{\citenamefont {Lein}\ \emph {et~al.}(2002)\citenamefont {Lein},
  \citenamefont {Hay}, \citenamefont {Velotta}, \citenamefont {Marangos},\ and\
  \citenamefont {Knight}}]{Lein2002}%
  \BibitemOpen
  \bibfield  {author} {\bibinfo {author} {\bibfnamefont {M.}~\bibnamefont
  {Lein}}, \bibinfo {author} {\bibfnamefont {N.}~\bibnamefont {Hay}}, \bibinfo
  {author} {\bibfnamefont {R.}~\bibnamefont {Velotta}}, \bibinfo {author}
  {\bibfnamefont {J.~P.}\ \bibnamefont {Marangos}},\ and\ \bibinfo {author}
  {\bibfnamefont {P.~L.}\ \bibnamefont {Knight}},\ }\bibfield  {title}
  {\bibinfo {title} {Interference effects in high-order harmonic generation
  with molecules},\ }\href {https://doi.org/10.1103/PhysRevA.66.023805}
  {\bibfield  {journal} {\bibinfo  {journal} {Phys. Rev. A}\ }\textbf {\bibinfo
  {volume} {66}},\ \bibinfo {pages} {023805} (\bibinfo {year}
  {2002})}\BibitemShut {NoStop}%
\bibitem [{\citenamefont {Torres}\ \emph {et~al.}(2007)\citenamefont {Torres},
  \citenamefont {Kajumba}, \citenamefont {Underwood}, \citenamefont {Robinson},
  \citenamefont {Baker}, \citenamefont {Tisch}, \citenamefont {de~Nalda},
  \citenamefont {Bryan}, \citenamefont {Velotta}, \citenamefont {Altucci},
  \citenamefont {Turcu},\ and\ \citenamefont {Marangos}}]{Torres2007}%
  \BibitemOpen
  \bibfield  {author} {\bibinfo {author} {\bibfnamefont {R.}~\bibnamefont
  {Torres}}, \bibinfo {author} {\bibfnamefont {N.}~\bibnamefont {Kajumba}},
  \bibinfo {author} {\bibfnamefont {J.~G.}\ \bibnamefont {Underwood}}, \bibinfo
  {author} {\bibfnamefont {J.~S.}\ \bibnamefont {Robinson}}, \bibinfo {author}
  {\bibfnamefont {S.}~\bibnamefont {Baker}}, \bibinfo {author} {\bibfnamefont
  {J.~W.~G.}\ \bibnamefont {Tisch}}, \bibinfo {author} {\bibfnamefont
  {R.}~\bibnamefont {de~Nalda}}, \bibinfo {author} {\bibfnamefont {W.~A.}\
  \bibnamefont {Bryan}}, \bibinfo {author} {\bibfnamefont {R.}~\bibnamefont
  {Velotta}}, \bibinfo {author} {\bibfnamefont {C.}~\bibnamefont {Altucci}},
  \bibinfo {author} {\bibfnamefont {I.~C.~E.}\ \bibnamefont {Turcu}},\ and\
  \bibinfo {author} {\bibfnamefont {J.~P.}\ \bibnamefont {Marangos}},\
  }\bibfield  {title} {\bibinfo {title} {Probing orbital structure of
  polyatomic molecules by high-order harmonic generation},\ }\href
  {https://doi.org/10.1103/PhysRevLett.98.203007} {\bibfield  {journal}
  {\bibinfo  {journal} {Phys. Rev. Lett.}\ }\textbf {\bibinfo {volume} {98}},\
  \bibinfo {pages} {203007} (\bibinfo {year} {2007})}\BibitemShut {NoStop}%
\bibitem [{\citenamefont {Li}\ \emph {et~al.}(2008)\citenamefont {Li},
  \citenamefont {Zhou}, \citenamefont {Lock}, \citenamefont {Patchkovskii},
  \citenamefont {Stolow}, \citenamefont {Kapteyn},\ and\ \citenamefont
  {Murnane}}]{Li2008}%
  \BibitemOpen
  \bibfield  {author} {\bibinfo {author} {\bibfnamefont {W.}~\bibnamefont
  {Li}}, \bibinfo {author} {\bibfnamefont {X.}~\bibnamefont {Zhou}}, \bibinfo
  {author} {\bibfnamefont {R.}~\bibnamefont {Lock}}, \bibinfo {author}
  {\bibfnamefont {S.}~\bibnamefont {Patchkovskii}}, \bibinfo {author}
  {\bibfnamefont {A.}~\bibnamefont {Stolow}}, \bibinfo {author} {\bibfnamefont
  {H.~C.}\ \bibnamefont {Kapteyn}},\ and\ \bibinfo {author} {\bibfnamefont
  {M.~M.}\ \bibnamefont {Murnane}},\ }\bibfield  {title} {\bibinfo {title}
  {Time-resolved dynamics in ${{N}}_2{{O}}_4$ probed using high harmonic
  generation},\ }\href {https://doi.org/10.1126/science.1163077} {\bibfield
  {journal} {\bibinfo  {journal} {Science}\ }\textbf {\bibinfo {volume}
  {322}},\ \bibinfo {pages} {1207} (\bibinfo {year} {2008})}\BibitemShut
  {NoStop}%
\bibitem [{\citenamefont {Lein}\ and\ \citenamefont {Rost}(2003)}]{Lein2003}%
  \BibitemOpen
  \bibfield  {author} {\bibinfo {author} {\bibfnamefont {M.}~\bibnamefont
  {Lein}}\ and\ \bibinfo {author} {\bibfnamefont {J.~M.}\ \bibnamefont
  {Rost}},\ }\bibfield  {title} {\bibinfo {title} {Ultrahigh harmonics from
  laser-assisted ion-atom collisions},\ }\href
  {https://doi.org/10.1103/PhysRevLett.91.243901} {\bibfield  {journal}
  {\bibinfo  {journal} {Phys. Rev. Lett.}\ }\textbf {\bibinfo {volume} {91}},\
  \bibinfo {pages} {243901} (\bibinfo {year} {2003})}\BibitemShut {NoStop}%
\bibitem [{\citenamefont {Schubert}\ \emph {et~al.}(2014)\citenamefont
  {Schubert}, \citenamefont {Hohenleutner}, \citenamefont {Langer},
  \citenamefont {Urbanek}, \citenamefont {Lange}, \citenamefont {Huttner},
  \citenamefont {Golde}, \citenamefont {Meier}, \citenamefont {Kira},
  \citenamefont {Koch},\ and\ \citenamefont {Huber}}]{Schubert2014}%
  \BibitemOpen
  \bibfield  {author} {\bibinfo {author} {\bibfnamefont {O.}~\bibnamefont
  {Schubert}}, \bibinfo {author} {\bibfnamefont {M.}~\bibnamefont
  {Hohenleutner}}, \bibinfo {author} {\bibfnamefont {F.}~\bibnamefont
  {Langer}}, \bibinfo {author} {\bibfnamefont {B.}~\bibnamefont {Urbanek}},
  \bibinfo {author} {\bibfnamefont {C.}~\bibnamefont {Lange}}, \bibinfo
  {author} {\bibfnamefont {U.}~\bibnamefont {Huttner}}, \bibinfo {author}
  {\bibfnamefont {D.}~\bibnamefont {Golde}}, \bibinfo {author} {\bibfnamefont
  {T.}~\bibnamefont {Meier}}, \bibinfo {author} {\bibfnamefont
  {M.}~\bibnamefont {Kira}}, \bibinfo {author} {\bibfnamefont {S.~W.}\
  \bibnamefont {Koch}},\ and\ \bibinfo {author} {\bibfnamefont
  {R.}~\bibnamefont {Huber}},\ }\bibfield  {title} {\bibinfo {title} {Sub-cycle
  control of terahertz high-harmonic generation by dynamical {{B}}loch
  oscillations},\ }\href {https://doi.org/10.1038/nphoton.2013.349} {\bibfield
  {journal} {\bibinfo  {journal} {Nature Photonics}\ }\textbf {\bibinfo
  {volume} {8}},\ \bibinfo {pages} {119} (\bibinfo {year} {2014})}\BibitemShut
  {NoStop}%
\bibitem [{\citenamefont {Corkum}(1993)}]{Corkum1993}%
  \BibitemOpen
  \bibfield  {author} {\bibinfo {author} {\bibfnamefont {P.~B.}\ \bibnamefont
  {Corkum}},\ }\bibfield  {title} {\bibinfo {title} {Plasma perspective on
  strong field multiphoton ionization},\ }\href
  {https://doi.org/10.1103/PhysRevLett.71.1994} {\bibfield  {journal} {\bibinfo
   {journal} {Phys. Rev. Lett.}\ }\textbf {\bibinfo {volume} {71}},\ \bibinfo
  {pages} {1994} (\bibinfo {year} {1993})}\BibitemShut {NoStop}%
\bibitem [{\citenamefont {Lewenstein}\ \emph {et~al.}(1994)\citenamefont
  {Lewenstein}, \citenamefont {Balcou}, \citenamefont {Ivanov}, \citenamefont
  {L'Huillier},\ and\ \citenamefont {Corkum}}]{Lewenstein1994}%
  \BibitemOpen
  \bibfield  {author} {\bibinfo {author} {\bibfnamefont {M.}~\bibnamefont
  {Lewenstein}}, \bibinfo {author} {\bibfnamefont {P.}~\bibnamefont {Balcou}},
  \bibinfo {author} {\bibfnamefont {M.~Y.}\ \bibnamefont {Ivanov}}, \bibinfo
  {author} {\bibfnamefont {A.}~\bibnamefont {L'Huillier}},\ and\ \bibinfo
  {author} {\bibfnamefont {P.~B.}\ \bibnamefont {Corkum}},\ }\bibfield  {title}
  {\bibinfo {title} {Theory of high-harmonic generation by low-frequency laser
  fields},\ }\href {https://doi.org/10.1103/PhysRevA.49.2117} {\bibfield
  {journal} {\bibinfo  {journal} {Phys. Rev. A}\ }\textbf {\bibinfo {volume}
  {49}},\ \bibinfo {pages} {2117} (\bibinfo {year} {1994})}\BibitemShut
  {NoStop}%
\bibitem [{\citenamefont {Ghimire}\ \emph {et~al.}(2011)\citenamefont
  {Ghimire}, \citenamefont {DiChiara}, \citenamefont {Sistrunk}, \citenamefont
  {Agostini}, \citenamefont {DiMauro},\ and\ \citenamefont
  {Reis}}]{Ghimire2011}%
  \BibitemOpen
  \bibfield  {author} {\bibinfo {author} {\bibfnamefont {S.}~\bibnamefont
  {Ghimire}}, \bibinfo {author} {\bibfnamefont {A.~D.}\ \bibnamefont
  {DiChiara}}, \bibinfo {author} {\bibfnamefont {E.}~\bibnamefont {Sistrunk}},
  \bibinfo {author} {\bibfnamefont {P.}~\bibnamefont {Agostini}}, \bibinfo
  {author} {\bibfnamefont {L.~F.}\ \bibnamefont {DiMauro}},\ and\ \bibinfo
  {author} {\bibfnamefont {D.~A.}\ \bibnamefont {Reis}},\ }\bibfield  {title}
  {\bibinfo {title} {Observation of high-order harmonic generation in a bulk
  crystal},\ }\href {https://doi.org/10.1038/nphys1847} {\bibfield  {journal}
  {\bibinfo  {journal} {Nature Physics}\ }\textbf {\bibinfo {volume} {7}},\
  \bibinfo {pages} {138} (\bibinfo {year} {2011})}\BibitemShut {NoStop}%
\bibitem [{\citenamefont {Vampa}\ and\ \citenamefont
  {Brabec}(2017)}]{Vampa2017}%
  \BibitemOpen
  \bibfield  {author} {\bibinfo {author} {\bibfnamefont {G.}~\bibnamefont
  {Vampa}}\ and\ \bibinfo {author} {\bibfnamefont {T.}~\bibnamefont {Brabec}},\
  }\bibfield  {title} {\bibinfo {title} {Merge of high harmonic generation from
  gases and solids and its implications for attosecond science},\ }\href
  {https://doi.org/10.1088/1361-6455/aa528d} {\bibfield  {journal} {\bibinfo
  {journal} {J. Phys. B}\ }\textbf {\bibinfo {volume} {50}},\ \bibinfo {pages}
  {083001} (\bibinfo {year} {2017})}\BibitemShut {NoStop}%
\bibitem [{\citenamefont {Kruchinin}\ \emph {et~al.}(2018)\citenamefont
  {Kruchinin}, \citenamefont {Krausz},\ and\ \citenamefont
  {Yakovlev}}]{Kruchinin2018}%
  \BibitemOpen
  \bibfield  {author} {\bibinfo {author} {\bibfnamefont {S.~Y.}\ \bibnamefont
  {Kruchinin}}, \bibinfo {author} {\bibfnamefont {F.}~\bibnamefont {Krausz}},\
  and\ \bibinfo {author} {\bibfnamefont {V.~S.}\ \bibnamefont {Yakovlev}},\
  }\bibfield  {title} {\bibinfo {title} {Colloquium: Strong-field phenomena in
  periodic systems},\ }\href {https://doi.org/10.1103/RevModPhys.90.021002}
  {\bibfield  {journal} {\bibinfo  {journal} {Rev. Mod. Phys.}\ }\textbf
  {\bibinfo {volume} {90}},\ \bibinfo {pages} {021002} (\bibinfo {year}
  {2018})}\BibitemShut {NoStop}%
\bibitem [{\citenamefont {Ghimire}\ and\ \citenamefont
  {Reis}(2019)}]{Ghimire2019}%
  \BibitemOpen
  \bibfield  {author} {\bibinfo {author} {\bibfnamefont {S.}~\bibnamefont
  {Ghimire}}\ and\ \bibinfo {author} {\bibfnamefont {D.~A.}\ \bibnamefont
  {Reis}},\ }\bibfield  {title} {\bibinfo {title} {High-harmonic generation
  from solids},\ }\href {https://doi.org/10.1038/s41567-018-0315-5} {\bibfield
  {journal} {\bibinfo  {journal} {Nature Physics}\ }\textbf {\bibinfo {volume}
  {15}},\ \bibinfo {pages} {10} (\bibinfo {year} {2019})}\BibitemShut {NoStop}%
\bibitem [{\citenamefont {Yue}\ and\ \citenamefont {Gaarde}(2021)}]{Yue2021}%
  \BibitemOpen
  \bibfield  {author} {\bibinfo {author} {\bibfnamefont {L.}~\bibnamefont
  {Yue}}\ and\ \bibinfo {author} {\bibfnamefont {M.~B.}\ \bibnamefont
  {Gaarde}},\ }\href@noop {} {\bibinfo {title} {Introduction to theory of
  high-harmonic generation in solids: tutorial}} (\bibinfo {year} {2021}),\
  \Eprint {https://arxiv.org/abs/2111.08669} {arXiv:2111.08669
  [physics.atom-ph]} \BibitemShut {NoStop}%
\bibitem [{\citenamefont {Golde}\ \emph {et~al.}(2008)\citenamefont {Golde},
  \citenamefont {Meier},\ and\ \citenamefont {Koch}}]{Golde2008}%
  \BibitemOpen
  \bibfield  {author} {\bibinfo {author} {\bibfnamefont {D.}~\bibnamefont
  {Golde}}, \bibinfo {author} {\bibfnamefont {T.}~\bibnamefont {Meier}},\ and\
  \bibinfo {author} {\bibfnamefont {S.~W.}\ \bibnamefont {Koch}},\ }\bibfield
  {title} {\bibinfo {title} {High harmonics generated in semiconductor
  nanostructures by the coupled dynamics of optical inter- and intraband
  excitations},\ }\href {https://doi.org/10.1103/PhysRevB.77.075330} {\bibfield
   {journal} {\bibinfo  {journal} {Phys. Rev. B}\ }\textbf {\bibinfo {volume}
  {77}},\ \bibinfo {pages} {075330} (\bibinfo {year} {2008})}\BibitemShut
  {NoStop}%
\bibitem [{\citenamefont {Vampa}\ \emph {et~al.}(2014)\citenamefont {Vampa},
  \citenamefont {McDonald}, \citenamefont {Orlando}, \citenamefont {Klug},
  \citenamefont {Corkum},\ and\ \citenamefont {Brabec}}]{Vampa2014}%
  \BibitemOpen
  \bibfield  {author} {\bibinfo {author} {\bibfnamefont {G.}~\bibnamefont
  {Vampa}}, \bibinfo {author} {\bibfnamefont {C.~R.}\ \bibnamefont {McDonald}},
  \bibinfo {author} {\bibfnamefont {G.}~\bibnamefont {Orlando}}, \bibinfo
  {author} {\bibfnamefont {D.~D.}\ \bibnamefont {Klug}}, \bibinfo {author}
  {\bibfnamefont {P.~B.}\ \bibnamefont {Corkum}},\ and\ \bibinfo {author}
  {\bibfnamefont {T.}~\bibnamefont {Brabec}},\ }\bibfield  {title} {\bibinfo
  {title} {Theoretical analysis of high-harmonic generation in solids},\ }\href
  {https://doi.org/10.1103/PhysRevLett.113.073901} {\bibfield  {journal}
  {\bibinfo  {journal} {Phys. Rev. Lett.}\ }\textbf {\bibinfo {volume} {113}},\
  \bibinfo {pages} {073901} (\bibinfo {year} {2014})}\BibitemShut {NoStop}%
\bibitem [{\citenamefont {Garg}\ \emph {et~al.}(2016)\citenamefont {Garg},
  \citenamefont {Zhan}, \citenamefont {Luu}, \citenamefont {Lakhotia},
  \citenamefont {Klostermann}, \citenamefont {Guggenmos},\ and\ \citenamefont
  {Goulielmakis}}]{Garg2016}%
  \BibitemOpen
  \bibfield  {author} {\bibinfo {author} {\bibfnamefont {M.}~\bibnamefont
  {Garg}}, \bibinfo {author} {\bibfnamefont {M.}~\bibnamefont {Zhan}}, \bibinfo
  {author} {\bibfnamefont {T.~T.}\ \bibnamefont {Luu}}, \bibinfo {author}
  {\bibfnamefont {H.}~\bibnamefont {Lakhotia}}, \bibinfo {author}
  {\bibfnamefont {T.}~\bibnamefont {Klostermann}}, \bibinfo {author}
  {\bibfnamefont {A.}~\bibnamefont {Guggenmos}},\ and\ \bibinfo {author}
  {\bibfnamefont {E.}~\bibnamefont {Goulielmakis}},\ }\bibfield  {title}
  {\bibinfo {title} {Multi-petahertz electronic metrology},\ }\href
  {https://doi.org/10.1038/nature19821} {\bibfield  {journal} {\bibinfo
  {journal} {Nature}\ }\textbf {\bibinfo {volume} {538}},\ \bibinfo {pages}
  {359} (\bibinfo {year} {2016})}\BibitemShut {NoStop}%
\bibitem [{\citenamefont {Luu}\ \emph {et~al.}(2015)\citenamefont {Luu},
  \citenamefont {Garg}, \citenamefont {Kruchinin}, \citenamefont {Moulet},
  \citenamefont {Hassan},\ and\ \citenamefont {Goulielmakis}}]{Luu2015}%
  \BibitemOpen
  \bibfield  {author} {\bibinfo {author} {\bibfnamefont {T.~T.}\ \bibnamefont
  {Luu}}, \bibinfo {author} {\bibfnamefont {M.}~\bibnamefont {Garg}}, \bibinfo
  {author} {\bibfnamefont {S.~Y.}\ \bibnamefont {Kruchinin}}, \bibinfo {author}
  {\bibfnamefont {A.}~\bibnamefont {Moulet}}, \bibinfo {author} {\bibfnamefont
  {M.~T.}\ \bibnamefont {Hassan}},\ and\ \bibinfo {author} {\bibfnamefont
  {E.}~\bibnamefont {Goulielmakis}},\ }\bibfield  {title} {\bibinfo {title}
  {Extreme ultraviolet high-harmonic spectroscopy of solids},\ }\href
  {https://doi.org/10.1038/nature14456} {\bibfield  {journal} {\bibinfo
  {journal} {Nature}\ }\textbf {\bibinfo {volume} {521}},\ \bibinfo {pages}
  {498} (\bibinfo {year} {2015})}\BibitemShut {NoStop}%
\bibitem [{\citenamefont {You}\ \emph {et~al.}(2017)\citenamefont {You},
  \citenamefont {Reis},\ and\ \citenamefont {Ghimire}}]{You2017}%
  \BibitemOpen
  \bibfield  {author} {\bibinfo {author} {\bibfnamefont {Y.~S.}\ \bibnamefont
  {You}}, \bibinfo {author} {\bibfnamefont {D.~A.}\ \bibnamefont {Reis}},\ and\
  \bibinfo {author} {\bibfnamefont {S.}~\bibnamefont {Ghimire}},\ }\bibfield
  {title} {\bibinfo {title} {Anisotropic high-harmonic generation in bulk
  crystals},\ }\href {https://doi.org/10.1038/nphys3955} {\bibfield  {journal}
  {\bibinfo  {journal} {Nature Physics}\ }\textbf {\bibinfo {volume} {13}},\
  \bibinfo {pages} {345} (\bibinfo {year} {2017})}\BibitemShut {NoStop}%
\bibitem [{\citenamefont {Kaneshima}\ \emph {et~al.}(2018)\citenamefont
  {Kaneshima}, \citenamefont {Shinohara}, \citenamefont {Takeuchi},
  \citenamefont {Ishii}, \citenamefont {Imasaka}, \citenamefont {Kaji},
  \citenamefont {Ashihara}, \citenamefont {Ishikawa},\ and\ \citenamefont
  {Itatani}}]{Kaneshima2018}%
  \BibitemOpen
  \bibfield  {author} {\bibinfo {author} {\bibfnamefont {K.}~\bibnamefont
  {Kaneshima}}, \bibinfo {author} {\bibfnamefont {Y.}~\bibnamefont
  {Shinohara}}, \bibinfo {author} {\bibfnamefont {K.}~\bibnamefont {Takeuchi}},
  \bibinfo {author} {\bibfnamefont {N.}~\bibnamefont {Ishii}}, \bibinfo
  {author} {\bibfnamefont {K.}~\bibnamefont {Imasaka}}, \bibinfo {author}
  {\bibfnamefont {T.}~\bibnamefont {Kaji}}, \bibinfo {author} {\bibfnamefont
  {S.}~\bibnamefont {Ashihara}}, \bibinfo {author} {\bibfnamefont {K.~L.}\
  \bibnamefont {Ishikawa}},\ and\ \bibinfo {author} {\bibfnamefont
  {J.}~\bibnamefont {Itatani}},\ }\bibfield  {title} {\bibinfo {title}
  {Polarization-resolved study of high harmonics from bulk semiconductors},\
  }\href {https://doi.org/10.1103/PhysRevLett.120.243903} {\bibfield  {journal}
  {\bibinfo  {journal} {Phys. Rev. Lett.}\ }\textbf {\bibinfo {volume} {120}},\
  \bibinfo {pages} {243903} (\bibinfo {year} {2018})}\BibitemShut {NoStop}%
\bibitem [{\citenamefont {Liu}\ \emph {et~al.}(2017)\citenamefont {Liu},
  \citenamefont {Li}, \citenamefont {You}, \citenamefont {Ghimire},
  \citenamefont {Heinz},\ and\ \citenamefont {Reis}}]{Liu2017}%
  \BibitemOpen
  \bibfield  {author} {\bibinfo {author} {\bibfnamefont {H.}~\bibnamefont
  {Liu}}, \bibinfo {author} {\bibfnamefont {Y.}~\bibnamefont {Li}}, \bibinfo
  {author} {\bibfnamefont {Y.~S.}\ \bibnamefont {You}}, \bibinfo {author}
  {\bibfnamefont {S.}~\bibnamefont {Ghimire}}, \bibinfo {author} {\bibfnamefont
  {T.~F.}\ \bibnamefont {Heinz}},\ and\ \bibinfo {author} {\bibfnamefont
  {D.~A.}\ \bibnamefont {Reis}},\ }\bibfield  {title} {\bibinfo {title}
  {High-harmonic generation from an atomically thin semiconductor},\ }\href
  {https://doi.org/10.1038/nphys3946} {\bibfield  {journal} {\bibinfo
  {journal} {Nature Physics}\ }\textbf {\bibinfo {volume} {13}},\ \bibinfo
  {pages} {262} (\bibinfo {year} {2017})}\BibitemShut {NoStop}%
\bibitem [{\citenamefont {Lakhotia}\ \emph {et~al.}(2020)\citenamefont
  {Lakhotia}, \citenamefont {Kim}, \citenamefont {Zhan}, \citenamefont {Hu},
  \citenamefont {Meng},\ and\ \citenamefont {Goulielmakis}}]{Lakhotia2020}%
  \BibitemOpen
  \bibfield  {author} {\bibinfo {author} {\bibfnamefont {H.}~\bibnamefont
  {Lakhotia}}, \bibinfo {author} {\bibfnamefont {H.~Y.}\ \bibnamefont {Kim}},
  \bibinfo {author} {\bibfnamefont {M.}~\bibnamefont {Zhan}}, \bibinfo {author}
  {\bibfnamefont {S.}~\bibnamefont {Hu}}, \bibinfo {author} {\bibfnamefont
  {S.}~\bibnamefont {Meng}},\ and\ \bibinfo {author} {\bibfnamefont
  {E.}~\bibnamefont {Goulielmakis}},\ }\bibfield  {title} {\bibinfo {title}
  {Laser picoscopy of valence electrons in solids},\ }\href
  {https://doi.org/10.1038/s41586-020-2429-z} {\bibfield  {journal} {\bibinfo
  {journal} {Nature}\ }\textbf {\bibinfo {volume} {583}},\ \bibinfo {pages}
  {55} (\bibinfo {year} {2020})}\BibitemShut {NoStop}%
\bibitem [{\citenamefont {Madsen}(2021)}]{Madsen2021}%
  \BibitemOpen
  \bibfield  {author} {\bibinfo {author} {\bibfnamefont {L.~B.}\ \bibnamefont
  {Madsen}},\ }\bibfield  {title} {\bibinfo {title} {Strong-field approximation
  for high-order harmonic generation in infrared laser pulses in the
  accelerated kramers-henneberger frame},\ }\href
  {https://doi.org/10.1103/PhysRevA.104.033117} {\bibfield  {journal} {\bibinfo
   {journal} {Phys. Rev. A}\ }\textbf {\bibinfo {volume} {104}},\ \bibinfo
  {pages} {033117} (\bibinfo {year} {2021})}\BibitemShut {NoStop}%
\bibitem [{\citenamefont {Yamada}\ and\ \citenamefont
  {Yabana}(2021)}]{Yamada2021}%
  \BibitemOpen
  \bibfield  {author} {\bibinfo {author} {\bibfnamefont {S.}~\bibnamefont
  {Yamada}}\ and\ \bibinfo {author} {\bibfnamefont {K.}~\bibnamefont
  {Yabana}},\ }\bibfield  {title} {\bibinfo {title} {Determining the optimum
  thickness for high harmonic generation from nanoscale thin films: An ab
  initio computational study},\ }\href
  {https://doi.org/10.1103/PhysRevB.103.155426} {\bibfield  {journal} {\bibinfo
   {journal} {Phys. Rev. B}\ }\textbf {\bibinfo {volume} {103}},\ \bibinfo
  {pages} {155426} (\bibinfo {year} {2021})}\BibitemShut {NoStop}%
\bibitem [{\citenamefont {Hansen}\ \emph {et~al.}(2017)\citenamefont {Hansen},
  \citenamefont {Deffge},\ and\ \citenamefont {Bauer}}]{PhysRevA.96.053418}%
  \BibitemOpen
  \bibfield  {author} {\bibinfo {author} {\bibfnamefont {K.~K.}\ \bibnamefont
  {Hansen}}, \bibinfo {author} {\bibfnamefont {T.}~\bibnamefont {Deffge}},\
  and\ \bibinfo {author} {\bibfnamefont {D.}~\bibnamefont {Bauer}},\ }\bibfield
   {title} {\bibinfo {title} {High-order harmonic generation in solid slabs
  beyond the single-active-electron approximation},\ }\href
  {https://doi.org/10.1103/PhysRevA.96.053418} {\bibfield  {journal} {\bibinfo
  {journal} {Phys. Rev. A}\ }\textbf {\bibinfo {volume} {96}},\ \bibinfo
  {pages} {053418} (\bibinfo {year} {2017})}\BibitemShut {NoStop}%
\bibitem [{\citenamefont {Tancogne-Dejean}\ \emph {et~al.}(2017)\citenamefont
  {Tancogne-Dejean}, \citenamefont {M\"ucke}, \citenamefont {K\"artner},\ and\
  \citenamefont {Rubio}}]{PhysRevLett.118.087403}%
  \BibitemOpen
  \bibfield  {author} {\bibinfo {author} {\bibfnamefont {N.}~\bibnamefont
  {Tancogne-Dejean}}, \bibinfo {author} {\bibfnamefont {O.~D.}\ \bibnamefont
  {M\"ucke}}, \bibinfo {author} {\bibfnamefont {F.~X.}\ \bibnamefont
  {K\"artner}},\ and\ \bibinfo {author} {\bibfnamefont {A.}~\bibnamefont
  {Rubio}},\ }\bibfield  {title} {\bibinfo {title} {Impact of the electronic
  band structure in high-harmonic generation spectra of solids},\ }\href
  {https://doi.org/10.1103/PhysRevLett.118.087403} {\bibfield  {journal}
  {\bibinfo  {journal} {Phys. Rev. Lett.}\ }\textbf {\bibinfo {volume} {118}},\
  \bibinfo {pages} {087403} (\bibinfo {year} {2017})}\BibitemShut {NoStop}%
\bibitem [{\citenamefont {Floss}\ \emph {et~al.}(2018)\citenamefont {Floss},
  \citenamefont {Lemell}, \citenamefont {Wachter}, \citenamefont {Smejkal},
  \citenamefont {Sato}, \citenamefont {Tong}, \citenamefont {Yabana},\ and\
  \citenamefont {Burgd\"orfer}}]{Floss2018}%
  \BibitemOpen
  \bibfield  {author} {\bibinfo {author} {\bibfnamefont {I.}~\bibnamefont
  {Floss}}, \bibinfo {author} {\bibfnamefont {C.}~\bibnamefont {Lemell}},
  \bibinfo {author} {\bibfnamefont {G.}~\bibnamefont {Wachter}}, \bibinfo
  {author} {\bibfnamefont {V.}~\bibnamefont {Smejkal}}, \bibinfo {author}
  {\bibfnamefont {S.~A.}\ \bibnamefont {Sato}}, \bibinfo {author}
  {\bibfnamefont {X.-M.}\ \bibnamefont {Tong}}, \bibinfo {author}
  {\bibfnamefont {K.}~\bibnamefont {Yabana}},\ and\ \bibinfo {author}
  {\bibfnamefont {J.}~\bibnamefont {Burgd\"orfer}},\ }\bibfield  {title}
  {\bibinfo {title} {Ab initio multiscale simulation of high-order harmonic
  generation in solids},\ }\href {https://doi.org/10.1103/PhysRevA.97.011401}
  {\bibfield  {journal} {\bibinfo  {journal} {Phys. Rev. A}\ }\textbf {\bibinfo
  {volume} {97}},\ \bibinfo {pages} {011401} (\bibinfo {year}
  {2018})}\BibitemShut {NoStop}%
\bibitem [{\citenamefont {Hansen}\ \emph {et~al.}(2018)\citenamefont {Hansen},
  \citenamefont {Bauer},\ and\ \citenamefont {Madsen}}]{Hansen2018}%
  \BibitemOpen
  \bibfield  {author} {\bibinfo {author} {\bibfnamefont {K.~K.}\ \bibnamefont
  {Hansen}}, \bibinfo {author} {\bibfnamefont {D.}~\bibnamefont {Bauer}},\ and\
  \bibinfo {author} {\bibfnamefont {L.~B.}\ \bibnamefont {Madsen}},\ }\bibfield
   {title} {\bibinfo {title} {Finite-system effects on high-order harmonic
  generation: From atoms to solids},\ }\href
  {https://doi.org/10.1103/PhysRevA.97.043424} {\bibfield  {journal} {\bibinfo
  {journal} {Phys. Rev. A}\ }\textbf {\bibinfo {volume} {97}},\ \bibinfo
  {pages} {043424} (\bibinfo {year} {2018})}\BibitemShut {NoStop}%
\bibitem [{\citenamefont {Bauer}\ and\ \citenamefont
  {Hansen}(2018)}]{Hansen2018_2}%
  \BibitemOpen
  \bibfield  {author} {\bibinfo {author} {\bibfnamefont {D.}~\bibnamefont
  {Bauer}}\ and\ \bibinfo {author} {\bibfnamefont {K.~K.}\ \bibnamefont
  {Hansen}},\ }\bibfield  {title} {\bibinfo {title} {High-harmonic generation
  in solids with and without topological edge states},\ }\href
  {https://doi.org/10.1103/PhysRevLett.120.177401} {\bibfield  {journal}
  {\bibinfo  {journal} {Phys. Rev. Lett.}\ }\textbf {\bibinfo {volume} {120}},\
  \bibinfo {pages} {177401} (\bibinfo {year} {2018})}\BibitemShut {NoStop}%
\bibitem [{\citenamefont {Yu}\ \emph {et~al.}(2021)\citenamefont {Yu},
  \citenamefont {Saalmann},\ and\ \citenamefont {Rost}}]{Chuan2021}%
  \BibitemOpen
  \bibfield  {author} {\bibinfo {author} {\bibfnamefont {C.}~\bibnamefont
  {Yu}}, \bibinfo {author} {\bibfnamefont {U.}~\bibnamefont {Saalmann}},\ and\
  \bibinfo {author} {\bibfnamefont {J.~M.}\ \bibnamefont {Rost}},\ }\href@noop
  {} {\bibinfo {title} {High harmonics from backscattering of delocalized
  electrons}} (\bibinfo {year} {2021}),\ \Eprint
  {https://arxiv.org/abs/2102.11208} {arXiv:2102.11208 [physics.atom-ph]}
  \BibitemShut {NoStop}%
\bibitem [{\citenamefont {Jensen}\ \emph {et~al.}(2021)\citenamefont {Jensen},
  \citenamefont {Iravani},\ and\ \citenamefont {Madsen}}]{Jensen2021}%
  \BibitemOpen
  \bibfield  {author} {\bibinfo {author} {\bibfnamefont {S.~V.~B.}\
  \bibnamefont {Jensen}}, \bibinfo {author} {\bibfnamefont {H.}~\bibnamefont
  {Iravani}},\ and\ \bibinfo {author} {\bibfnamefont {L.~B.}\ \bibnamefont
  {Madsen}},\ }\bibfield  {title} {\bibinfo {title} {Edge-state-induced
  correlation effects in two-color pump-probe high-order harmonic generation},\
  }\href {https://doi.org/10.1103/PhysRevA.103.053121} {\bibfield  {journal}
  {\bibinfo  {journal} {Phys. Rev. A}\ }\textbf {\bibinfo {volume} {103}},\
  \bibinfo {pages} {053121} (\bibinfo {year} {2021})}\BibitemShut {NoStop}%
\bibitem [{\citenamefont {Jensen}\ and\ \citenamefont
  {Madsen}(2021)}]{Jensen2021_2}%
  \BibitemOpen
  \bibfield  {author} {\bibinfo {author} {\bibfnamefont {S.~V.~B.}\
  \bibnamefont {Jensen}}\ and\ \bibinfo {author} {\bibfnamefont {L.~B.}\
  \bibnamefont {Madsen}},\ }\bibfield  {title} {\bibinfo {title} {Edge-state
  and bulklike laser-induced correlation effects in high-harmonic generation
  from a linear chain},\ }\href {https://doi.org/10.1103/PhysRevB.104.054309}
  {\bibfield  {journal} {\bibinfo  {journal} {Phys. Rev. B}\ }\textbf {\bibinfo
  {volume} {104}},\ \bibinfo {pages} {054309} (\bibinfo {year}
  {2021})}\BibitemShut {NoStop}%
\bibitem [{\citenamefont {Imai}\ \emph {et~al.}(2020)\citenamefont {Imai},
  \citenamefont {Ono},\ and\ \citenamefont
  {Ishihara}}]{PhysRevLett.124.157404}%
  \BibitemOpen
  \bibfield  {author} {\bibinfo {author} {\bibfnamefont {S.}~\bibnamefont
  {Imai}}, \bibinfo {author} {\bibfnamefont {A.}~\bibnamefont {Ono}},\ and\
  \bibinfo {author} {\bibfnamefont {S.}~\bibnamefont {Ishihara}},\ }\bibfield
  {title} {\bibinfo {title} {High harmonic generation in a correlated electron
  system},\ }\href {https://doi.org/10.1103/PhysRevLett.124.157404} {\bibfield
  {journal} {\bibinfo  {journal} {Phys. Rev. Lett.}\ }\textbf {\bibinfo
  {volume} {124}},\ \bibinfo {pages} {157404} (\bibinfo {year}
  {2020})}\BibitemShut {NoStop}%
\bibitem [{\citenamefont {Murakami}\ \emph {et~al.}(2018)\citenamefont
  {Murakami}, \citenamefont {Eckstein},\ and\ \citenamefont
  {Werner}}]{Murakami2018}%
  \BibitemOpen
  \bibfield  {author} {\bibinfo {author} {\bibfnamefont {Y.}~\bibnamefont
  {Murakami}}, \bibinfo {author} {\bibfnamefont {M.}~\bibnamefont {Eckstein}},\
  and\ \bibinfo {author} {\bibfnamefont {P.}~\bibnamefont {Werner}},\
  }\bibfield  {title} {\bibinfo {title} {High-harmonic generation in {{M}}ott
  insulators},\ }\href {https://doi.org/10.1103/PhysRevLett.121.057405}
  {\bibfield  {journal} {\bibinfo  {journal} {Phys. Rev. Lett.}\ }\textbf
  {\bibinfo {volume} {121}},\ \bibinfo {pages} {057405} (\bibinfo {year}
  {2018})}\BibitemShut {NoStop}%
\bibitem [{\citenamefont {Takayoshi}\ \emph {et~al.}(2019)\citenamefont
  {Takayoshi}, \citenamefont {Murakami},\ and\ \citenamefont
  {Werner}}]{Takayoshi2019}%
  \BibitemOpen
  \bibfield  {author} {\bibinfo {author} {\bibfnamefont {S.}~\bibnamefont
  {Takayoshi}}, \bibinfo {author} {\bibfnamefont {Y.}~\bibnamefont
  {Murakami}},\ and\ \bibinfo {author} {\bibfnamefont {P.}~\bibnamefont
  {Werner}},\ }\bibfield  {title} {\bibinfo {title} {High-harmonic generation
  in quantum spin systems},\ }\href
  {https://doi.org/10.1103/PhysRevB.99.184303} {\bibfield  {journal} {\bibinfo
  {journal} {Phys. Rev. B}\ }\textbf {\bibinfo {volume} {99}},\ \bibinfo
  {pages} {184303} (\bibinfo {year} {2019})}\BibitemShut {NoStop}%
\bibitem [{\citenamefont {Tancogne-Dejean}\ \emph {et~al.}(2018)\citenamefont
  {Tancogne-Dejean}, \citenamefont {Sentef},\ and\ \citenamefont
  {Rubio}}]{PhysRevLett.121.097402}%
  \BibitemOpen
  \bibfield  {author} {\bibinfo {author} {\bibfnamefont {N.}~\bibnamefont
  {Tancogne-Dejean}}, \bibinfo {author} {\bibfnamefont {M.~A.}\ \bibnamefont
  {Sentef}},\ and\ \bibinfo {author} {\bibfnamefont {A.}~\bibnamefont
  {Rubio}},\ }\bibfield  {title} {\bibinfo {title} {Ultrafast modification of
  {{H}}ubbard {{$U$}} in a strongly correlated material: Ab initio
  high-harmonic generation in {{N}}i{{O}}},\ }\href
  {https://doi.org/10.1103/PhysRevLett.121.097402} {\bibfield  {journal}
  {\bibinfo  {journal} {Phys. Rev. Lett.}\ }\textbf {\bibinfo {volume} {121}},\
  \bibinfo {pages} {097402} (\bibinfo {year} {2018})}\BibitemShut {NoStop}%
\bibitem [{\citenamefont {Murakami}\ \emph {et~al.}(2021)\citenamefont
  {Murakami}, \citenamefont {Takayoshi}, \citenamefont {Koga},\ and\
  \citenamefont {Werner}}]{Murakami2021}%
  \BibitemOpen
  \bibfield  {author} {\bibinfo {author} {\bibfnamefont {Y.}~\bibnamefont
  {Murakami}}, \bibinfo {author} {\bibfnamefont {S.}~\bibnamefont {Takayoshi}},
  \bibinfo {author} {\bibfnamefont {A.}~\bibnamefont {Koga}},\ and\ \bibinfo
  {author} {\bibfnamefont {P.}~\bibnamefont {Werner}},\ }\bibfield  {title}
  {\bibinfo {title} {High-harmonic generation in one-dimensional {{M}}ott
  insulators},\ }\href {https://doi.org/10.1103/PhysRevB.103.035110} {\bibfield
   {journal} {\bibinfo  {journal} {Phys. Rev. B}\ }\textbf {\bibinfo {volume}
  {103}},\ \bibinfo {pages} {035110} (\bibinfo {year} {2021})}\BibitemShut
  {NoStop}%
\bibitem [{\citenamefont {Lysne}\ \emph {et~al.}(2020)\citenamefont {Lysne},
  \citenamefont {Murakami},\ and\ \citenamefont {Werner}}]{Lysne2020}%
  \BibitemOpen
  \bibfield  {author} {\bibinfo {author} {\bibfnamefont {M.}~\bibnamefont
  {Lysne}}, \bibinfo {author} {\bibfnamefont {Y.}~\bibnamefont {Murakami}},\
  and\ \bibinfo {author} {\bibfnamefont {P.}~\bibnamefont {Werner}},\
  }\bibfield  {title} {\bibinfo {title} {Signatures of bosonic excitations in
  high-harmonic spectra of {{M}}ott insulators},\ }\href
  {https://doi.org/10.1103/PhysRevB.101.195139} {\bibfield  {journal} {\bibinfo
   {journal} {Phys. Rev. B}\ }\textbf {\bibinfo {volume} {101}},\ \bibinfo
  {pages} {195139} (\bibinfo {year} {2020})}\BibitemShut {NoStop}%
\bibitem [{\citenamefont {Essler}\ \emph {et~al.}(2005)\citenamefont {Essler},
  \citenamefont {Frahm}, \citenamefont {Göhmann}, \citenamefont {Klümper},\
  and\ \citenamefont {Korepin}}]{Hubbard}%
  \BibitemOpen
  \bibfield  {author} {\bibinfo {author} {\bibfnamefont {F.~H.~L.}\
  \bibnamefont {Essler}}, \bibinfo {author} {\bibfnamefont {H.}~\bibnamefont
  {Frahm}}, \bibinfo {author} {\bibfnamefont {F.}~\bibnamefont {Göhmann}},
  \bibinfo {author} {\bibfnamefont {A.}~\bibnamefont {Klümper}},\ and\
  \bibinfo {author} {\bibfnamefont {V.~E.}\ \bibnamefont {Korepin}},\ }\href
  {https://doi.org/10.1017/CBO9780511534843} {\emph {\bibinfo {title} {The
  One-Dimensional Hubbard Model}}}\ (\bibinfo  {publisher} {Cambridge
  University Press},\ \bibinfo {address} {Cambridge},\ \bibinfo {year}
  {2005})\BibitemShut {NoStop}%
\bibitem [{\citenamefont {Tomita}\ and\ \citenamefont
  {Nasu}(2001)}]{Tomita2001}%
  \BibitemOpen
  \bibfield  {author} {\bibinfo {author} {\bibfnamefont {N.}~\bibnamefont
  {Tomita}}\ and\ \bibinfo {author} {\bibfnamefont {K.}~\bibnamefont {Nasu}},\
  }\bibfield  {title} {\bibinfo {title} {Quantum fluctuation effects on light
  absorption spectra of the one-dimensional extended {{H}}ubbard model},\
  }\href {https://doi.org/10.1103/PhysRevB.63.085107} {\bibfield  {journal}
  {\bibinfo  {journal} {Phys. Rev. B}\ }\textbf {\bibinfo {volume} {63}},\
  \bibinfo {pages} {085107} (\bibinfo {year} {2001})}\BibitemShut {NoStop}%
\bibitem [{\citenamefont {Mahan}(2000)}]{Mahan}%
  \BibitemOpen
  \bibfield  {author} {\bibinfo {author} {\bibfnamefont {G.~D.}\ \bibnamefont
  {Mahan}},\ }\href@noop {} {\emph {\bibinfo {title} {Many-Particle Physics}}}\
  (\bibinfo  {publisher} {Kluwer Academic},\ \bibinfo {address} {New York},\
  \bibinfo {year} {2000})\ p.~\bibinfo {pages} {24}\BibitemShut {NoStop}%
\bibitem [{\citenamefont {Gorsso}\ and\ \citenamefont
  {Parravicini}(2014)}]{Solid_State_Physics}%
  \BibitemOpen
  \bibfield  {author} {\bibinfo {author} {\bibfnamefont {G.}~\bibnamefont
  {Gorsso}}\ and\ \bibinfo {author} {\bibfnamefont {G.~P.}\ \bibnamefont
  {Parravicini}},\ }\href@noop {} {\emph {\bibinfo {title} {Solid State
  Physics}}}\ (\bibinfo  {publisher} {Academic Press},\ \bibinfo {address}
  {Cambridge, Massachusetts},\ \bibinfo {year} {2014})\ p.~\bibinfo {pages}
  {26}\BibitemShut {NoStop}%
\end{thebibliography}
\end{document}